\documentclass[a4 paper, 12pt]{article}
\usepackage[utf8]{inputenc}
\usepackage[T1]{fontenc}
\usepackage{graphicx}
\usepackage[frozencache,cachedir=.]{minted}
\usepackage{grffile}
\usepackage{longtable}
\usepackage{wrapfig}
\usepackage{rotating}
\usepackage{multirow}
\usepackage[normalem]{ulem}
\usepackage{amsmath}
\usepackage{textcomp}
\usepackage{amssymb}
\usepackage{capt-of}
\usepackage{hyperref}
\usepackage{amsmath,amsfonts,amsthm,amssymb}
\usepackage{graphicx}
\usepackage{xfrac}
\usepackage{complexity}
\usepackage[inner=2.5cm,outer=2.5cm,top=2.5cm,bottom=2.5cm]{geometry}

\newtheorem{theorem}{Theorem}

\usepackage{enumitem}
\usepackage{tikz}
\usetikzlibrary{graphs}
\usetikzlibrary{calc}

\setenumerate[1]{itemsep=2pt,partopsep=0pt,parsep=\parskip,topsep=3pt}
\setitemize[1]{itemsep=2pt,partopsep=0pt,parsep=\parskip,topsep=3pt}
\setdescription[1]{itemsep=2pt,partopsep=0pt,parsep=\parskip,topsep=3pt}

\author{Jiatu Li\footnote{lijt19@mails.tsinghua.edu.cn}\\\small{Institute for Interdisciplinary Information Sciences, Tsinghua University}}
\date{}
\title{Formalization of PAL$\cdot$S5 in Proof Assistant}
\hypersetup{
 pdfauthor={Li Jiatu},
 pdftitle={Formalization of PAL in Proof Assistant},
 pdflang={English}}
\begin{document}
\maketitle
\begin{abstract}
  As an experiment to the application of proof assistant for logic research, we formalize the model and proof system for multi-agent modal logic S5 with PAL-style dynamic modality in Lean theorem prover. We provide a formal proof for the reduction axiom of public announcement, and the soundness and completeness of modal logic S5, which can be typechecked with Lean 3.19.0. The complete proof is now available at Github.
\end{abstract}

\section{Introduction}

\textit{Proof assistant} is a useful tool to organize and check formal proofs, which can be used to develop reliable softwares and formal mathematical proofs. Despite of the impressive stories including the verification of the \textit{independence of continuum hypothesis} \cite{ch20} and the \textit{four-color theorem} \cite{fct-coq05}, it is still hopeless for common mathematicians to use proof assistant for their daily work, since a formal proof is neither concise nor readable for human. However, it may be handy for those scenarios that only require specific types of mathematics, for example, in developing logic systems. In this work, we formalize PAL (a simple dynamic epistemic logic) in Lean theorem prover, as an experiment to the application of proof assistant for logicians. 

Public Announcement Logic (PAL) is a simple but widely used dynamic epistemic logic which can model information update in multi-agent scenario. We provide a formal proof for reduction axioms of public announcement and soundness and completeness of modal logic S5, in Lean theorem prover. The complete proof has about 1600 lines of code, which is available at Github\footnote{https://github.com/ljt12138/Formalization-PAL} and is typechecked with Lean 3.19.0.

We formalize (1) the syntax and semantics of PAL$\cdot$S5, (2) the recursion theorem for PAL updates, (3) a Hilbert-style proof system for S5 and (4) the soundness and completeness for S5. The following two theorems are the main goal of our formalization.
\begin{theorem}[Recursion Theorem]\label{thm:recursion}
  For any PAL sentence $\varphi$, there exists a sentence $\psi$ without public announcement that is logically equivalent to $\varphi$, i.e. share the same evaluation in all models. 
\end{theorem}
\begin{theorem}[Completeness Theorem]\label{thm:completeness}
  The proof system for S5 (without public announcement) is complete.  
\end{theorem}

\subsection{Dynamic Epistemic Logic}

Dynamic epistemic logic (DEL for short) is an active area of logic that aims to combine information update with knowledge, belief and other epistemic notions based on modal logic. DEL has wide applications on the analysis of information update in multi-agent scenarios, for example, games and social choice. Technically, dynamic epistemic logic extends a modal logic system (representing epistemic abilities of agents) with dynamic modalities (whose truth value is defined by model update).

\subsection{Lean Theorem Prover}

Lean is a high-order theorem prover with type theory based framework, usually called Calculus of Inductively Constructions (CIC) \cite{lean15,pfenning89}. It natively supports intuitionist logic, while classical reasoning can be naturally introduced with extra axioms in standard library. Inductive definition is natively supported in CIC, which is extensively used to define syntax of language and proof system. The mathematical library \texttt{mathlib}, which is still in developing by Lean community, supports basic notions including relation and its properties used to define semantics.

\subsection{Outline}

Throughout this paper we shall not present details of proof code in Lean for simplicity. Instead, we will show the definitions of significant results and explain the idea of proof. We will also mention the main technical problems in the proof. Interested readers are encouraged to see the code on Github for the formal proof.

In Section \ref{sec:pre} we provide a brief introduction to public announcement logic and the standard workflow to deal with dynamic epistemic logic. In Section \ref{sec:definition} we present the definition of syntax, semantics and proof system for PAL$\cdot$S5 in our formalization. In Section \ref{sec:recursion} and \ref{sec:completeness}, we describe the proof of Theorem \ref{thm:recursion} and \ref{thm:completeness}, respectively. Finally we discuss related work, provide our comments on the formal proof and present the conclusion in the last two sections.

\section{Preliminaries}
\label{sec:pre}

\subsection{Intuition of Dynamic Epistemic Logic}

Informally, dynamic epistemic logic aims to model the following sentences.
\begin{enumerate}
\item I \textit{believe} that today will be sunny, but I \textit{don't know} whether it is true.
\item \textit{After} his grade is claimed, Eric \textit{knows} that his parents will be unhappy.
\item \textit{After} Alice knows that Bob does not know the answer, she \textit{believes} that she has a high rank in the exam.
\item $\dots$
\end{enumerate}
Here, a DEL sentence involves agents (I, Alice, Bob, etc), epistemic relations (knowledge, belief, etc) and information update (public announcement, private communication, etc). 

Epistemic relations may have some logical properties, for example, if I \textit{know} $\varphi$, then $\varphi$ must be true. To realize the internal structure of dynamic epistemic logic, we need to define the \textit{model} for it, and construct a \textit{complete proof system}.   

\subsection{Public Announcement Logic}

Public announcement logic is a special case of DEL, in which \textit{public announcement} is the only supported type of information update. A PAL sentence is defined as $\varphi := p\mid\perp\mid\varphi\to\varphi\mid\Box_i\varphi\mid[!\varphi]\varphi$. Intuitively, $\Box_i\varphi$ means that the agent $i$ \textit{knows} $\varphi$, while $[!\varphi]\psi$ means that $\psi$ holds after $\varphi$ is announced. PAL can be used to analysis games with information update, for example, Muddy Children puzzle\footnote{Folklore, see \url{https://plato.stanford.edu/entries/dynamic-epistemic/} for details.}.

The semantics of PAL is defined by standard model of modal logic and information update. The model is a tuple $(\mathcal{W}, \alpha, \mathcal{A}, f, \{\sim_i\}|_{i\in\alpha})$, where $\mathcal{W},\alpha$ and $\mathcal{A}$ is the set of worlds, atomic propositions and agents, respectively. The function $f:\mathcal{W}\times\alpha\to\{\mathrm{True},\mathrm{False}\}$ assign truth value for atomic propositions at every world, and the binary relations $\sim_i$ denote the epistemic ability of agent $i$. In modal logic S5, $\sim_i$ should be an equivalence relation, which does not necessarily hold in weaker systems like K, T, S4 and B.

Now we define the evaluation of sentences in models. Suppose $M=(\mathcal{W}, \alpha, \mathcal{A}, f, \{\sim_i\}|_{i\in\alpha})$ is a model and $s\in\mathcal{W}$, we can define the evaluation of $\varphi$ in $M$ at $s$ as follow.                                         %
\begin{align*}                                                                                                                                                                                                           %
  (M, s) \models p &\mathbf{\ \ \ iff\ \ \ } f(s, p) = \mathrm{True} \\                                                                                                                                                  %
  (M, s) \models \varphi\to\psi &\mathbf{\ \ \ iff\ \ \ } \mathrm{if\ } (M, s)\models\varphi \mathrm{,\ then\ } (M, s)\models\psi \\                                                                                     %
  (M, s) \models \Box_i\varphi &\mathbf{\ \ \ iff\ \ \ } \mathrm{forall\ }t\sim_i s, (M, t)\models\varphi \\                                                                                                             %
  (M, s) \models [!\varphi]\psi & \mathbf{\ \ \ iff\ \ \ } \mathrm{if\ }(M, s)\models\varphi\mathrm{,\ then\ }(M|_{\varphi}, s)\models\psi.                                                                              %
\end{align*}                                                                                                                                                                                                             %
Note that here $M|_{\varphi}$ denote the model obtained from $M$ by restricting to those worlds that $\varphi$ holds, or equivalently, by removing all the epistemic links from the worlds that $\varphi$ does not hold. Also the contradiction $\perp$ does not hold in all models, i.e. $(M, s)\not\models\perp$ for all $M$ and $s$. 

\subsection{Standard Workflow for DEL}

Suppose a model theoretical interpretation of a DEL is given, constructing a complete proof system is usually done via a standard procedure consisting of two steps: (1) find a procedure to reduce dynamic modalities in sentences, and (2) construct a sound and complete proof system for static sentences. We shall introduce such workflow using PAL as an example.
\begin{description}
\item[Reducing Dynamicality] Let $\varphi$ be a PAL sentence, which may involve dynamic modalities (i.e. public announcement modalities $[!\varphi]$) and epistemic modalities (i.e. the knowledge modality $\Box$). Usually the update of model can be embedded into the sentences, for example, one may verify that
  \[
    ([!\varphi](\psi\to\gamma))\leftrightarrow([!\varphi]\psi\to[!\varphi]\gamma). 
  \]
  These rules are usually called \textit{reduction axioms}. By recursively applying reduction axioms one may eliminate all dynamic modalities in a sentence, which is what Theorem \ref{thm:recursion} states.  
\item[Static Proof System] After reducing all dynamic modalities, we only need to deal with sentences consisting of $\Box$ and Boolean connectives. A complete proof system is usually found by adding proper axioms and rules corresponding to modalities to the \textit{minimum modal logic}, whose axioms are propositional tautologies and the modal distribution axiom $\Box(\varphi\to\psi)\to\varphi\to\psi$. 
\end{description}

\subsection{Type Theory, Proof and Mathematics}

Like many proof assistant, the underlaying theory of Lean is \textit{Calculus of Inductively Constructions} \cite{pfenning89} based on \textit{type theory}. Type theory studies the property of type systems for programming languages, which is related to (intuitionist) proof systems\footnote{This means that in Lean, to support classical reasoning, additional axioms need to be introduced. From a programmer's perspective, additional axioms can be considered as external libraries.} by the well-known \textit{Curry-Howard Correspondence}.

Curry-Howard Correspondence roughly means that the relationship between propositions and proofs is similar to the relationship between types and programs. For example, the Modus Ponens rule of intuitionist propositional logic
\[
(\mathrm{MP})\frac{\Gamma\vdash\varphi\to\psi\ \ \ \Gamma\vdash\varphi}{\Gamma\vdash\psi}
\]
may be read in two ways.
\begin{enumerate}
\item Let $\Gamma$ be the set of assumptions. If both $\varphi\to\psi$ and $\varphi$ are provable from the assumptions, then $\psi$ is also provable.  
\item Let $\Gamma$ be the set of type information. If there is a function from $\varphi$ to $\psi$ and an element of type $\varphi$, then evaluating the function at the element obtains a term of type $\psi$.  
\end{enumerate}

This idea motivates us to consider propositions as a special kind of type, so that proving a theorem is done by giving a program of such type. In Lean (and many other proof assistants like Coq), one can provide a proof in two different ways (or both).
\begin{enumerate}
\item Directly construct the proof (program) in a functional programming language.
\item Prove with \textit{tactics}: the programmer may interactively do a goal-oriented proof, which is similar to the usual way to present an informal proof. 
\end{enumerate}

The correspondence between type system and proof system makes it a promising meta-language for mathematics. A type may be considered as a collection of objects, but unlike a set, it does not have many strong properties like the \textit{Axiom of Choice}. It can be shown that set theory with arbitrarily large cardinality can be modeled in Lean, so that it is a ``universal'' formal system. 

\subsection{Inductive Definition}

Many finite mathematical objects are essentially defined by induction, for example, natural numbers, trees, finite lists and logic sentences. In Lean, it is natively supported to construct a type by inductive definition. For instance, the natural numbers can be defined as follow.
\begin{minted}[escapeinside=@@,mathescape=true]{lean}
inductive nat : Type
| O : nat          -- $0$ is a nat
| S : nat → nat    -- if $n$ is a nat, (S $n$) is also a nat
\end{minted}
Here \texttt{O} and \texttt{S} are called the \textit{type constructors}, and this definition means that \texttt{nat} is the \textit{minimal} type that is closed under these type constructors.

The mechanics behind inductive definition is simple: the system will define the type constructors and automatically add some additional axioms to make it the minimal type under the induction rules. For example, the following three axioms are frequently used to deal with inductively defintion types. 
\begin{enumerate}
\item For an inductively defined type $\alpha$ with type constructors $f_1,f_2,\dots,f_k$, there is an axiom to ensure that every instance $n$ of type $\alpha$ is constructed by one of the constructors. This axiom ensures that we can do \textit{case analysis} on an inductively defined type.
\item For an inductively defined type $\alpha$ and its type constructor $f_i$, there is an axiom to ensure that if $n_1$ and $n_2$ are constructed by $f_i$ with different parameters, then they are not equal to each other. This axiom ensures that we can distinguish the instances constructed by different parameters.
\item For an inductively defined type $\alpha$, there is an axiom for \textit{structural induction} on this type. That is, for any predicate $\mathcal{P}$ over $\alpha$, if every constructor $f_i$ of $\alpha$ satisfies that for all parameters $m_1,m_2,\dots,m_t$,
  \[
    (\forall i, (m_i:\alpha)\to\mathcal{P}(m_i))\Rightarrow \mathcal{P}(f_i(m_1,m_2,\dots,m_t)),
  \]
  then $\mathcal{P}$ holds for all instances of type $\alpha$. 
\end{enumerate}

Note that we don't need to apply these axioms directly; these axioms have been natively supported, such that we can perform case analysis or structural induction over inductively defined types with high-level tactics like \texttt{induction} or \texttt{case}. 

\section{Definitions of PAL}\label{sec:definition}

In this section we present the definition of syntax and semantics for PAL$\cdot$S5, and a complete Hilbert-style proof system of S5 in Lean. 

\subsection{Syntax}

A PAL sentence is defined as $\varphi := p\mid\perp\mid\varphi\to\varphi\mid\Box_i\varphi\mid[!\varphi]\varphi$, which is naturally translated to a formal inductive definition as following. In this definition, $\alpha$ is the type for propositional letters and \texttt{agent} is the type for agents. 

\begin{minted}[escapeinside=@@,mathescape=true]{lean}
inductive sentence (@$\alpha$@ : Type) (agent : Type) : Type 
| atom_prop : @$\alpha$@ → sentence                          -- atomic propositions
| perp : sentence                                   -- contradiction $\perp$
| imply : sentence → sentence → sentence            -- imply $\varphi\to\psi$ 
| box : agent → sentence → sentence                 -- box $\Box_i\varphi$
| announce : sentence → sentence → sentence         -- announce $[!\varphi]\psi$
\end{minted}

Notations like $\Box$ and $\Diamond$ are defined for convenience, where $\Diamond:=\lnot\Box\lnot$ is the existential modality of $\Box$. Other propositional connectivities are defined in terms of $\to$ and $\perp$, for example, $\lnot\varphi$ is defined as $\varphi\to\perp$.

\subsection{Semantics}

The model of PAL is defined by a \textit{structure} called worlds, containing an evaluation map, the epistemic links for each agent, and a proof that the epistemic link is an equivalence relation. Structure in Lean defines a type class; this definition means that any type equipped by the evaluation map, the epistemic links and the equivalence proof is a correct type for worlds.  

\begin{minted}[escapeinside=@@,mathescape=true]{lean}
structure worlds (@$\alpha$@ agent W : Type) : Type := 
(f : W → @$\alpha$@ → Prop)
(view : agent → W → W → Prop)
(equiv : @$\forall$@(a : agent), equivalence (view a))
\end{minted}

The evaluation of sentences in worlds $M$ at world $s$ can be defined recursively. For instance, we define $\Box_i\varphi$ holds at $(M, s)$, if for all $s\sim_i t$, $\varphi$ holds at $(M, t)$. In Lean this definition is expressed by the case in the recursive definition of \texttt{evaluate}.
\mint[escapeinside=@@,mathescape=true]{lean}{| (M, s) @$\Box$@(i : @$\varphi$@) := @$\forall$@ t : W, M.view i s t → evaluate (M, t) @$\varphi$@}

In order to define the evaluation of public announcement modality we need to define the restriction $M|_{\varphi}$ of model. We generalize the definition of restriction to arbitrary predicate $\mathcal{P}$ over sentences, denoted by $M|_{\mathcal{P}}$. Usual restriction $M_{\varphi}$ is the special case when
\[
  \mathcal{P}(s)\iff (M, s)\models\varphi.
\]
We use such definition for technical convenience: if $M|_{\varphi}$ is defined direction, then restriction and evaluate are defined by mutual recursion, hence additional well-founded proof will be required.  

The definition of restriction of worlds $M|_{\mathcal{P}}$ is slightly different from the traditional version. In order to remove $\lnot\mathcal{P}$ worlds from the type $W$, we need to define a new type $W'$ and adapt the epistemic links for it, which is not convenient. We use the alternative definition: we remove epistemic links with an endpoint at a $\lnot\mathcal{P}$ world, so that all $\lnot\mathcal{P}$ worlds are isolated\footnote{Note that node removal may be essential for global existential modality. In such case, we may extend our model by a predicate $\mathcal{Q}$ identifying all remained worlds; we only need to update the predicate instead of removing elements from the type of worlds.}. Then we prove that the epistemic links remain to be equivalence relations, hence $M|_{\mathcal{P}}$ is indeed an instance of \texttt{worlds}. 

Finally we define some notions that are useful latter. A sentence $\varphi$ is a \textit{tautology}, or $\vDash\varphi$, if it holds in all models; two sentences $\varphi$ and $\psi$ are \textit{logically equivalent}, if they share the same truth value in all models. These definitions are easy to implement in Lean. 

\subsection{Proof System}

We also present a complete Hilbert-style proof system for S5, which is sufficient for PAL since the public announcement modalities can be eliminated via Theorem \ref{thm:recursion}. For simplicity, we define a proof system only for one sentence without set of assumptions. The axioms and rules are listed in the Appendix; and some redundant axioms are introduced for convenience.

We implement our proof system with an inductively defined type \texttt{proof} for the syntax of proof terms, and an inductively defined property \texttt{proof\_of} to define how the proof terms work. For example, we define $\varphi$ is provable, or $\vdash\varphi$, as follow.
\begin{minted}[escapeinside=@@,mathescape=true]{lean}
def provable {@$\alpha$@ agent : Type} (@$\varphi$@ : sentence @$\alpha$@ agent) :=
  @$\exists$@ pf : proof @$\alpha$@ agent, proof_of pf @$\varphi$@
\end{minted}
Here we separate the syntax and semantics of proof system, since it is convenience for further development of structural proof theory for it.  

It is easy to see that if a correct proof term \texttt{pf} is given, showing that it is correct can be easily done by applying appropriate rules of \texttt{proof\_of}. With the feature of meta-programming, we define a tactic \texttt{prover} to do that automatically. In order to prove a concrete proposition in our proof system, we only need to provide proof term and call \texttt{prover} to verify it. The following piece of code demonstrates how to prove $\varphi\to\varphi$ in our proof system.
\begin{minted}[escapeinside=@@,mathescape=true]{lean}
lemma id_provable {@$\alpha$@ agent : Type} (@$\varphi$@ : sentence @$\alpha$@ agent) :
  @$\vdash\varphi\rightarrowtail\varphi$@ :=
begin
  existsi (proof.mp (proof.mp (proof.ax2 _ _ _)
                              (proof.ax1 _ _))
                    (proof.ax1 @$\varphi\ \varphi$@)),
  prover
end
\end{minted}
The tactic \texttt{existsi} provide the proof term for the proposition $\varphi\to\varphi$, whose correctness is automatically verified by our tactic \texttt{prover}.

Note that sometimes\footnote{This relies on the ability of \textit{type inference} of Lean. However, type inference of CIC (or even weaker systems) is \textit{undecidable}, hence the heuristic algorithm may not give correct answer. This is why we still need to fill $\varphi$ into one of the axiom schemes.} we may use the axioms schemes without identifying how it is substituted: the placeholders ``\texttt{\_}'' in the proofs are automatically filled with proper types by the system. 

\section{Recursion Theorem}\label{sec:recursion}

In this section we describe the formal proof of Theorem \ref{thm:recursion}. Recall that two sentences are called logically equivalent, denoted by $\varphi\equiv\psi$, if they share the same truth value in all models. To express this theorem, we inductively define a sentence $\varphi$ is \textit{static} if it does not involve public announcement modality. The definition of the theorem in our formalization is the following.

\begin{minted}[escapeinside=@@,mathescape=true]{lean}
theorem reduction_dynamics {@$\alpha$@ agent : Type} (@$\varphi$@ : sentence @$\alpha$@ agent) :
  @$\exists\psi,\varphi\equiv\psi\ \land$@ static @$\psi$@
\end{minted}

The proof of recursion theorem requires two steps. Firstly, we verify that the following reduction axioms\footnote{Note that the reduction axiom for $\Box$ is slightly different from commonly used version $[!\varphi]\Box_i\psi\equiv\varphi \to \Box_i(\varphi\to[!\varphi]\psi)$; one may carefully check that our reduction axiom is also valid.} for public announcement modality are sound.
\begin{equation}
\begin{matrix}
  [!\varphi]p & \equiv & \varphi\to p \\
  [!\varphi]\perp & \equiv & \varphi\to\perp \\
  [!\varphi]\psi\to\gamma & \equiv & [!\varphi]\psi \to [!\varphi]\gamma \\
  [!\varphi]\Box_i\psi & \equiv & \varphi \to \Box_i[!\varphi]\psi
\end{matrix}
\end{equation}
Then, we preform structural induction on $\varphi$, to show that all the public announcement modalities involved can be eliminated by applying reduction axioms in appropriate order. 

\subsection{Verifying Reduction Axioms}

Verifying reduction axioms for public announcement modality is straightforward. We explain the reduction axiom for implication $[!\varphi]\psi\to\gamma\equiv[!\varphi]\psi \to [!\varphi]\gamma$ as an example.
\begin{minted}[escapeinside=@@,mathescape=true]{lean}
lemma recursion_imply {@$\alpha$@ agent : Type} : @$\forall$@ (@$\varphi\ \psi\ \gamma$@ : sentence @$\alpha$@ agent),
  @$([!\varphi](\psi\rightarrowtail\gamma)) \equiv ([!\varphi]\psi) \rightarrowtail ([!\varphi]\gamma)$@
\end{minted}

After unfolding the definition of logical equivalent, we need to show that for any model $(M, s)$, these two sentences share the same truth value. We unfold the evaluation of both sides: the left hand side means that after removing the epistemic links connecting to $\lnot\varphi$ worlds, if $s$ is a $\psi$ world, then it is a $\gamma$ world; the right hand side means that if $s$ is a $\psi$ world after removing the epistemic links, then it is a $\gamma$ world after removing epistemic links. Clearly these two conditions are equivalent. 

Note that the reduction axiom for $\Box$ modality is a little bit tedious, since we need to alter current world during evaluation. We may need to perform a careful case study with the property that the epistemic link is an equivalence relation. 

\subsection{Structural Induction}

Now we show that any sentence $\varphi$ can be reduced to a logically equivalent one without public announcement. Induction on the structure of $\varphi$. If $\varphi=p$ or $\perp$, the result is trivial. Assume that $\varphi=\psi\to\gamma$, we use the induction hypothesis to obtain static sentences $\psi'$ and $\gamma'$, such that $\psi'\equiv\psi$ and $\gamma'\equiv\gamma$; then it's easy to verify that $\psi'\to\gamma'\equiv\psi\to\gamma$ and is static. The same method also works for $\varphi=\Box_i\psi$.

The only situation left is that $\varphi=[!\psi]\gamma$, from induction hypothesis we may assume that $\psi$ and $\gamma$ are static. The following lemma gives the remaining part of the proof.
\begin{minted}[escapeinside=@@,mathescape=true]{lean}
lemma reduction_lemma {@$\alpha$@ agent : Type} (@$\varphi\ \psi$@ : sentence @$\alpha$@ agent) :
  static @$\varphi$@ → static @$\psi$@ → @$\exists\gamma$@, @$([!\varphi]\psi) \equiv \gamma\ \land$@ static @$\gamma$@
\end{minted}
That is, if $\varphi$ and $\psi$ are both static, $[!\varphi]\psi$ can be reduced to a logically equivalent sentence that is static. This can be proved by structural induction on $\psi$. Since it is static, it can only be a atomic proposition, a contradiction, an implication or a $\Box$ sentence, which is exactly where we use four reduction axioms for public announcement modalities.

Note that some properties of logically equivalence is required for this proof; interested readers are recommended to read the code in \texttt{dynamic.lean} at Github. 

\section{Completeness Theorem}\label{sec:completeness}

In this section we describe the formal proof of Theorem \ref{thm:completeness}. Recall that a proof system is called \textit{sound} if every provable sentence is a tautology; it is called \textit{complete} if every tautology is provable. Soundness of our proof system is quite easy to verify which is given in Appendix, but the completeness is non-trivial.

Our proof follows Chapter 5 of \cite{johan2010}, with slight modification to make it work for type-theory based formalization. The high-level idea of the proof goes as follow.
\begin{enumerate}
\item Show that any consistent set $\Gamma$ can be extended to a maximal consistent set.
\item Show that all maximal consistent sets forms a model $M$ with appropriate definition of epistemic links. 
\item Show that if we define the evaluation map as $f(\Gamma,p)=\mathrm{True}$ if and only if $p\in\Gamma$, then $\varphi\in\Gamma$ if and only if $\varphi$ holds at $(M, \Gamma)$.
\item Show that a consistent sentence is satisfiable, which implies the completeness.  
\end{enumerate}
The following five subsections describe the definitions and the details of the four steps above. 

\subsection{Definition of Consistent Sets}

In order to formalize the result we firstly need to define consistent set and maximal consistent set. 

We define a \texttt{decider} to be a property over sentences, which can be considered as a set containing all sentences satisfying this property. A decider is a \textit{consistent set} if every finite subset of it does not imply contradiction; and a decider is a \textit{maximal consistent set} (or \texttt{consistent\_decider} in our code) if it is a consistent set, and for every sentence $\varphi$, either $\varphi$ or $\lnot\varphi$ is inside it. For convenience we pack a decider and the proof that it is a maximal consistent set together, to form a type called \texttt{decider\_world}, defined as follow. We also define the relation \texttt{decides} for \texttt{decider\_world} over sentences.
\begin{minted}[escapeinside=@@,mathescape=true]{lean}
inductive decider_world (@$\alpha$@ agent : Type) : Type
| mk (d : decider @$\alpha$@ agent) : consistent_decider d → decider_world
def decides {@$\alpha$@ agent : Type}
    : decider_world @$\alpha$@ agent → sentence @$\alpha$@ agent → Prop
| (decider_world.mk d h) @$\varphi$@ := d @$\varphi$@
\end{minted}

\subsection{Complete Extension via Lindenbaum Construction}

In the first step of the proof, we need to show that any consistent set can be extended to a maximal consistent. This result is defined as follow.

\begin{minted}[escapeinside=@@,mathescape=true]{lean}
theorem dcomplete_extension {@$\alpha$@ agent : Type} [encodable (sentence @$\alpha$@ agent)] :
   @$\forall$@ d : decider @$\alpha$@ agent, consistent_set d →
     @$\exists$@ d' : decider_world @$\alpha$@ agent, (@$\forall\varphi$@, d @$\varphi$@ → decides d' @$\varphi$@)
\end{minted}

Note that here we assume that the sentences are \textit{encodable}, i.e. there is an injective mapping from sentences into $\mathbb N$. We use the \textit{Lindenbaum construction} to extend a consistent set to a maximal one. Let $\Gamma_0$ be the original consistent set, we define $\Gamma_n$ for all $n$ recursively as follow.
\begin{enumerate}
\item If $n$ is not the encoding of any sentence, $\Gamma_n = \Gamma_{n-1}$.
\item Otherwise let $\varphi$ be the sentence with encoding $n$. If $\Gamma_n = \Gamma_{n-1}\cup\{\varphi\}$; otherwise $\Gamma_n=\Gamma_{n-1}\cup\{\lnot\varphi\}$. 
\end{enumerate}
We need to verify that every $\Gamma_n$ is consistent, and $\Gamma_n\subseteq\Gamma_m$ if $n\le m$. Moreover, $\Gamma=\cup_n\Gamma_n$ is also consistent. Since either $\varphi$ or $\lnot\varphi$ is contained in $\Gamma_n$ where $n$ is the encoding of $\varphi$, we know that $\Gamma$ is actually maximal, which completes the constructions. 

\subsection{Henkin Model}

In the second step of the proof, we use \textit{Henkin model} to connect the syntactic representation (i.e. maximal consistent sets) of modal logic with its semantics. Let $W$ be the type of all consistent deciders, we define the epistemic link $\sim_i$ by the following relation \texttt{access}.
\begin{minted}[escapeinside=@@,mathescape=true]{lean}
def access {@$\alpha$@ agent : Type} (i : agent) (s t : decider_world @$\alpha$@ agent) :=
  @$\forall\varphi$@ : sentence @$\alpha$@ agent, decides s @$\Box$@(i:@$\varphi$@) → decides t @$\varphi$@
\end{minted}
That is, $s\sim_i t$ if and only if $\varphi\in t$ for any $\Box_i\varphi\in s$. It is easy to verify that $\sim_i$ is indeed an equivalence relation, so that we can create an instance of the structure \texttt{henkin\_worlds}, meaning that the type of all consistent deciders with the epistemic link defined above is actually a model. Note that here is the only place to use the axiom of reflexivity, transitivity and symmetry in the entire completeness proof.
\begin{minted}[escapeinside=@@,mathescape=true]{lean}
def henkin_worlds (@$\alpha$@ agent : Type) :
  worlds @$\alpha$@ agent (decider_world @$\alpha$@ agent) :=
{
  f := ...,            -- define the evaluation map
  view := ...,         -- define the access relation
  equiv := ...         -- prove that $\mathtt{view}$ is equivalent
}
\end{minted}

\subsection{Correctness of Henkin Model}

Let $M$ be the Henkin model defined above and $s\in W$ be a consistent decider. In the third step of the proof we need to show that, if the evaluation map $f(s,p)$ is defined as $p\in s$ for all proposition letter $p$, then a static sentence is valid at $(M, s)$ if and only if it is contained in $s$. The formal definition of the theorem is stated as follow.
\begin{minted}[escapeinside=@@,mathescape=true]{lean}
theorem henkin_correctness {@$\alpha$@ agent : Type} [encodable (sentence @$\alpha$@ agent)]:
  @$\forall\varphi$@, static @$\varphi$@ → @$\forall$@ s : decider_world @$\alpha$@ agent,
    (decides s @$\varphi$@ @$\leftrightarrow$@ (henkin_worlds @$\alpha$@ agent, s) @$\vDash$@ @$\varphi$@)
\end{minted}

The proof of the theorem is done by structural induction on $\varphi$. If $\varphi$ is an atomic proposition or a contradiction then the proof is straightforward. The case when $\varphi=\psi\to\gamma$ is tedious but intuitive, which is the same as the completeness proof for propositional logic.

The hardest part is the case $\varphi=\Box_i\psi$; we need to show that if $\Box_i\psi$ is true at $(M, s)$, $\Box_i\psi$ must be contained in $s$. The idea is as follow. Towards a contradiction assume that $\Box_i\psi\notin s$, by the maximality we can see that $\lnot\Box_i\psi\in s$. Consider whether $t=\{\lnot\psi\}\cup\{\gamma\mid\Box_i\gamma\in s\}$ is consistent.
\begin{enumerate}
\item If $t$ is consistent, then by theorem \texttt{dcomplete\_extension}, it can be extended to a consistent decider. By the definition of $\sim_i$ we can see that $s\sim_i t$. Since $\lnot\psi\in t$, by induction hypothesis, $\lnot\psi$ holds at $(M, t)$, a contradiction to the fact that $\Box_i\psi$ holds at $(M, s)$.
\item Otherwise, it is possible to deduce contradiction from a finite subset $\Gamma$ of $t$. If $\lnot\psi$ is not inside $\Gamma$, $s$ cannot be consistent, a contradiction; otherwise we can see that there exists a finite conjunction of $\gamma_k$ with $\Box_i\gamma_k\in s$ denoted by $\gamma^*$, such that $\gamma^*\to\psi$ is provable. Then by truth rule and modal distributive axiom (see Appendix), $\Box_i\gamma^*\to\Box_i\psi$ is provable. Since $\Box_i\gamma_k\in s$ for any $k$, it's easy to verify that $\Box_i\psi\in s$, with the fact that $s$ is close under modus ponens due to maximality and consistency. 
\end{enumerate}
This case study is done by lemma \texttt{neighborhood} in \texttt{henkin\_model} in our formalization.  

\subsection{Completeness Theorem}

The last part of the proof is to combine all the results above. Suppose that $\varphi$ is a tautology, towards a contradiction that $\varphi$ is not provable, we can see that $\lnot\varphi$ must be consistent. Therefore if $\lnot\varphi$ is satisfiable, there must be a model in which $\varphi$ does not hold, contradiction to the fact that $\varphi$ is a tautology. Hence we only need to show that the consistent set $\{\lnot\varphi\}$ is satisfiable. This is actually straightforward: by Lindenbaum construction it is possible to obtain a consistent decider $s$ containing $\lnot\varphi$, then by the correctness of Henkin model, $\lnot\varphi$ holds at $(M, s)$, which exactly means that $\lnot\varphi$ is satisfiable.  

\begin{minted}[escapeinside=@@,mathescape=true]{lean}
theorem completeness {@$\alpha$@ agent : Type} [encodable (sentence @$\alpha$@ agent)]
  (@$\varphi$@ : sentence @$\alpha$@ agent) : static @$\varphi$@ → @$\vDash$@ @$\varphi$@ → @$\vdash$@ @$\varphi$@ :=
begin
  intros st h, classical, by_contra,
  have h@$_1$@ : consistent_set (@$\lambda$@ @$\psi$@, @$\psi$@ = @$\varphi\rightarrowtail\perp$@),
  { ... },
  cases dcomplete_extension (@$\lambda$@ @$\psi$@, @$\psi$@ = @$\varphi\rightarrowtail\perp$@) h@$_1$@ with s h@$_2$@,
  have st' : static (@$\varphi\rightarrowtail\perp$@),
  { ... },
  have h@$_2$@ : henkin_model s @$\vDash$@ @$\varphi\rightarrowtail\perp$@,
  { ... },
  simp at h@$_2$@, apply h@$_2$@, apply h
end
\end{minted}

The formal proof of this part is shown above, which is quite readable. We firstly show that $\{\lnot\varphi\}$ is consistent, then with Lindenbaum construction (theorem \texttt{dcomplete\_extension}), we obtain a consistent decider containing $\lnot\varphi$. Finally we conclude that $\lnot\varphi$ holds at $(M, s)$ which leads to a contradiction. 

\section{Related Work}

Formalizing logic systems and their meta-theorems has been done for different logic systems in different theorem provers. For instance, \cite{prop15} formalizes the completeness theorem of propositional logic for natural deduction system in Agda; \cite{coq20} presents the completeness proof of first-order logic in Coq standard library.

The completeness of modal logic S5 has also been formalized in \cite{bentzen19} in Lean theorem prover, which is similar to our completeness proof. Our approach differs from it by choice of proof system and supporting of multi-agent scenario. We have not notice other formalization for dynamic epistemic logic (for example, for belief revision, probability and dependency) in theorem provers.  

\section{Conclusion}

In this work we present a computer-checked formalization of public announcement logic with recursion theorem for public announcement modality and completeness theorem for modal logic S5. From this experiment one may see that proof assistant is sufficient to model logic concepts like language, models and proofs.

\subsection{Challenges}

A sequence of work remains to be done for our ultimate goal, that is to build a suitable system for logicians to define logic systems, interactively develop proof systems and prove other meta-theorems. From our point of view, there are two major challenges. 

\textbf{The trade-off between expressiveness and convenience.} Proof assistant is a general-purpose solution for formal verifications, which is expressive but not convenient. Softwares like SAT solvers or model checking program are easy to use, but they can only be applied to specific tasks about logic systems. Clearly a suitable toolkit for logic should be both expressive and convenient: commonly used functions (like developing a complete proof system) should be natively supported, while it should be possible extend it for more specific applications. In such case, developing frameworks on proof assistants could be an ideal way to balance expressiveness and convenience. 

\textbf{More automations.} Automation is critical for real-world formal verifications; without automation tools, even inequalities of natural numbers can be hard to prove. One may dream that with well-designed algorithms we only need to write down the theorems without proving them. Unfortunately, general-purpose automated theorem proving has been a hard problem in AI for decades. In order to reduce human aid, we need to design a lot of domain specific automation algorithm for developing logic systems.  

\subsection{Further Work}

We shall point out some possible directions for further work.

\begin{description}
\item[Extend to other dynamic epistemic logic.] Dynamic epistemic logic is an active research field in modern logic community; besides public announcement logic, theory for belief revision, probability, dependency and other notion has been developed based on modal logic \cite{johan2016}. It is possible to extend our formalization to more kinds of dynamic logic.   
\item[Extend to other proof systems.] In this formalization, we consider a simple Hilbert-style system, which is not convenient for reasoning. For further development we may need to implement more complicated systems like sequent calculus or natural deduction; we may implement a ``compiler'' from our simple system to stronger systems, which can naturally translate the completeness result to those systems.
\item[Formalize proof theory.] Besides the completeness, proof theory is also critical to realize logic systems. For example, one may wonder whether the axioms and rules are redundant. Formalizing standard methods like \textit{cut elimination theorem} is also an interesting problem.
\end{description}

\paragraph{Acknowledgement} We are thankful to Johan van Benthem and Fenrong Liu for valuable comments on the formalization of DEL and many other points in this work.

\bibliographystyle{alpha}  
\bibliography{paper2}   

\newpage
\appendix

\section{The Proof System and Soundness}

Our proof system consists of ten axioms and seven rules; only seven axioms and two rules are necessary, the others are added for convenience. Note that all the redundant axioms and rules are about propositional logic, hence the soundness of redundant rules is easy to verify. 

\begin{table}[htb]
  \centering
\begin{tabular}{|l|l|l|}
\hline
\multirow{7}{*}{Necessary Axioms} & S1      & $\varphi\to\psi\to\varphi$                                                                                                                                         \\ \cline{2-3} 
                                  & S2      & $(\varphi\to\psi\to\gamma)\to(\varphi\to\psi)\to\varphi\to\gamma$                                                                                                  \\ \cline{2-3} 
                                  & S3      & $\lnot\lnot\varphi\to\varphi$                                                                                                                                      \\ \cline{2-3} 
                                  & Distr   & $\Box_i(\varphi\to\psi)\to\Box_i\varphi\to\Box_i\psi$                                                                                                              \\ \cline{2-3} 
                                  & Ref     & $\Box_i\varphi\to\varphi$                                                                                                                                          \\ \cline{2-3} 
                                  & Trans   & $\Box_i\Box_i\varphi\to\Box_i\varphi$                                                                                                                              \\ \cline{2-3} 
                                  & Sym     & $\lnot\Box_i\varphi\to\Box_i(\lnot\Box_i\varphi)$                                                                                                                  \\ \hline
\multirow{2}{*}{Necessary Rules}  & MP      & if $\vdash\varphi\to\psi$ and $\vdash\varphi$, then $\vdash\psi$                                                                                                   \\ \cline{2-3} 
                                  & Truth   & if $\vdash\varphi$, then $\vdash\Box_i\varphi$                                                                                                                     \\ \hline
\multirow{3}{*}{Redundant Axioms} & S2'     & $\varphi\to\lnot\lnot\varphi$                                                                                                                                      \\ \cline{2-3} 
                                  & S3'     & $(\varphi\to\psi)\to(\lnot\varphi\to\psi)\to\psi$                                                                                                                  \\ \cline{2-3} 
                                  & Conjmp  & $(\varphi\to\psi)\land\varphi\to\psi$                                                                                                                            \\ \hline
\multirow{5}{*}{Redundant Rules}  & Conj    & if $\displaystyle\vdash\bigwedge_{i\in\Gamma_1}\varphi_i\to\psi$ and $\Gamma_1\subseteq\Gamma_2$, then $\displaystyle\vdash\bigwedge_{i\in\Gamma_2}\varphi\to\psi$ \\ \cline{2-3} 
                                  & Conjl   & if $\vdash\varphi\to\gamma$, then $\vdash\varphi\land\psi\to\gamma$                                                                                                \\ \cline{2-3} 
                                  & Conjr   & if $\vdash\psi\to\gamma$, then $\vdash\varphi\land\psi\to\gamma$                                                                                                   \\ \cline{2-3} 
                                  & Uncurry & if $\vdash\varphi\to\psi\to\gamma$, then $\vdash\varphi\land\psi\to\gamma$                                                                                         \\ \cline{2-3} 
                                  & Curry   & if $\vdash\varphi\land\psi\to\gamma$, then $\vdash\varphi\to\psi\to\gamma$                                                                                         \\ \hline
\end{tabular}
\caption{List of axioms and rules.}
\end{table}

In fact, redundant axioms and rules are added into the system during development of the completeness proof, instead of designed at the beginning.

The soundness proof is quite simple with automation of Lean; soundness for most rules and axioms is proved simply by unfolding the definition and call automation tactic \texttt{tauto} for logical reasoning. The only rule that requires non-trivial human aid is \texttt{Conj}, which is also the most complicated rule in our system intuitively. 

\end{document}